\documentclass[%
 preprint,
 amsmath,amssymb,
]{revtex4-1}

\usepackage{graphicx}
\usepackage{amssymb}
\usepackage{setspace}
\usepackage[utf8]{inputenc}
\usepackage{array}
\usepackage{natbib}
\usepackage{amsmath}
\usepackage{multirow}
\usepackage{hyperref}

\usepackage{color}
\definecolor{redcolor}{rgb}{1.0,0.,0.}

\begin{document}

\preprint{}

\title{A comparative analysis of knowledge acquisition performance in complex networks}

\author{Lucas Guerreiro$^1$,  Filipi N. Silva$^2$ and Diego R. Amancio$^1$}

\affiliation{$^1$Institute of Mathematics and Computer Science, University of S\~ao Paulo, S\~ao Carlos, Brazil\\
$^2$Indiana University Network Science Institute, Bloomington, Indiana 47408, USA\\
}

\date{\today}

\begin{abstract}
Discovery processes have been an important topic in the network science field. The exploration of nodes can be understood as the knowledge acquisition process taking place in the network, where nodes represent concepts and edges are the semantical relationships between concepts. While some studies have analyzed the performance of the knowledge acquisition process in particular network topologies, here we performed a systematic performance analysis in well-known dynamics and topologies. Several interesting results have been found. Overall, all learning curves displayed the same learning shape, with different speed rates. We also found ambiguities in the feature space describing the learning curves, meaning that the same knowledge acquisition curve can be generated in different combinations of network topology and dynamics. A surprising example of such patterns are the learning curves obtained from random and Waxman networks: despite the very distinct characteristics in terms of global structure, several curves from different models turned out to be similar. All in all, our results suggest that different learning strategies can lead to the same learning performance. From the network reconstruction point of view, however, this means that learning curves of observed sequences should be combined with other sequence features if one aims at inferring network topology from observed sequences.
\end{abstract}

\maketitle



\section{Introduction}
Many real-world systems can be naturally represented by sequences corresponding to chains of events or transitions between states, including human actions~\cite{barbosa2018human}, machine  workflow~\cite{sipser1996introduction}, scientists mobility~\cite{franzoni2012foreign} and language~\cite{correa2020semantic}. Communication can also be accomplished by encoding and decoding data into sequences of symbols or continuous signals. Indeed, a significant portion of datasets derived from real-world systems is available in this form. For a complex system, one can understand that sequences can be generated by a process driving the changes among states across a certain space of allowed transitions~\cite{ARRUDA2019}.

Network science has been employed to represent a great variety of complex systems~\cite{costa2011analyzing,costa2007characterization,amancio2015topological,Mata2020,levnajic2013derivative,ban2017robust}. In recent studies, complex networks have displayed the potential to represent the space of transitions between states for many types of systems~\cite{correa2020semantic,Kim2016,Koponen2014}. In this context, the driving processes generating sequences are represented by stochastic walks of a variety of heuristics. An example of this case is the knowledge acquisition process~\cite{ARRUDA2017}, in which nodes represent knowledge that is connected according to how related they are. One or multiple agents (such as researchers) navigate in this knowledge space, which is unknown from the start, and discoveries are made when the agents visit new nodes. In such a system, sequences are derived by the paths taken by the agents.

While Markov chains~\cite{norris1998markov} are a simple way to model and recover the inherent network of transition probabilities, it relies on considering that the studied phenomenon is driven by a simple stochastic process with no \emph{a priori} knowledge of its space. Many real-systems, however, may present more intricate driving stochastic dynamics (which may depend on long term memory or properties of the nodes, for instance). An example of that system is urban transportation, where agents navigate across a system of roads with possibly predefined origin and destinations. The paths taken by connecting these endpoints cannot be driven solely based on local probabilities. Also, the inherent space of state transitions can display a variety of different topologies~\cite{costa2011analyzing} in contrast to more well-defined structures, such as regular graphs, as a consequence, even simple stochastic dynamics can lead to intricate sequences~\cite{ARRUDA2019}.

In many real-world problems, only the sequences generated by the system are observed. Thus, having a way to discriminate characteristics that are either consequence of the dynamics or from the network can lead to a better understanding of the studied phenomenon. A simple property derived from sequences that can be differently impacted by both of these aspects is the rate of appearance of new symbols. This corresponds to the exploration coverage of a network under the action of a walking dynamics, which is also related to the learning curve in a knowledge acquisition process. This property is also related to how well an agent performs in discovering knowledge.

To our knowledge, no previous study focused on a systematic analysis among the dynamics, networks, and the sequences generated by them. Here we analyzed the coverage curves for sequences obtained from four random walk dynamics and four network models with different topological structures. At first, we are interested in knowing if the coverage curves are already good criteria for determining both the model and the dynamics used to generate a sequence.

Our analysis revealed that, among the considered stochastic walk dynamics using only local network information, the \emph{true self-avoiding dynamics} (TSAW) was found to present the best performance in coverage rate for the considered network models.
In addition to that, different patterns for the performances of coverage rate were observed. Aside from TSAW, the ranking based on performance of exploration for different sets of walk dynamics tends to depend on the network structure. For instance, when the stochastic walk is biased according to the node degree, better performance is attained when the network is sparse and the walks are biased towards preferring highly connected nodes. On the other hand, if the network is denser, better performance is reached when the walk avoids highly connected nodes. We also encountered situations in which there exists ambiguity in the coverage property for certain combinations of dynamics and network models. This indicates that it would be possible to swap the dynamics and the inherent structure and even so, attain similar coverage curves.
These developments could shed a light on the analysis of the mechanisms leading to text generation, for instance, to better understand how the vocabulary grows along with the text.

The following section explores the related literature to the problem of modeling real-world phenomena in terms of networks, dynamics, and sequences. Next, the methodology is presented alongside the description of the considered network models and dynamics. Results are presented together with discussions, which is followed by conclusions.

\section{Related Works} \label{sec:related}
\label{S:2}

Random walks (RW) have been studied in many networked applications~\cite{barat1995statistics,meerschaert2006coupled,Comin2020,correa2017patterns}. In the early studies of the emergent network science field, the properties of RW was investigated in power-law distributed networks. In~\cite{Adamic2001}, the authors compared the efficiency of random and self-avoiding walks in transferring messages through the network. Hubs were found to play the role of centralizing and distributing information to other nodes. Most importantly, this finding revealed that the efficiency of discovering new nodes depends on the topology of the underlying network.

The process of network discovering has been approached by several recent studies~\cite{ARRUDA2019,lima2018dynamics,ARRUDA2017,Thagard1992,Herrero2019}. In~\cite{ARRUDA2017}, the authors compared the learning speed of several dynamics for particular network topologies. Specifically, they analyzed  how effective different dynamics are when discovering new nodes in the network. In addition to traditional random walks, this study considered also random walks with Lévy flights~\cite{sutantyo2010multi}. Thus, the agents were allowed to visit any node in the network in the next step with a certain probability. The authors found that more frequent jumps favors the discovery rate, specially in Barabási-Albert networks. In particular topologies, though, jumps were found not to be as effective. This is the case of geographic networks. Another interesting finding is that the discovery of new nodes occurs with different speed in different network regions. The core -- as identified via  accessibility (entropy diversity)~\cite{travenccolo2008accessibility,de2016using} -- tends to be covered faster than the network borders.

In~\cite{lima2018dynamics} the authors studied the efficiency of agents walking over the network to learn the structure of the network. Differently from other works, the authors considered a model where knowledge discovered by different agents is integrated in a specific entity of the system. This system is referred to as \emph{network brain}. This type of dynamics was intended to represent e.g. the knowledge acquisition when mapping communities of similar interests in the Web. The most surprising result arising from this study is the fact that the learning behavior, considering variations of the self-avoiding walk, has  a very weak dependence on the considered dynamics and network topologies.

The problem of knowledge acquisition in networks has also been studied in the context of information theory applications~\cite{ARRUDA2019}. In~\cite{ARRUDA2019}, distinct random walks are performed over different topologies. The sequence of visited nodes generates a sequence of symbols, which is further analyzed in function of the observed compression ratio -- computed via Huffman coding. Finally, such a sequence is used to reconstruct the original network, and the error is analyzed for distinct topologies and agent dynamics.  Several interesting results have been found using the framework combining knowledge acquisition and information theory. Interestingly, the best performance in the framework constructed for representing the phenomena of compression (during transmission) and reconstruction of networks revealed that a simple knitted network model~\cite{costa2007knitted} yielded the best performance. This finding is compatible with the idea that language is optimized for transmission~\cite{i2003least}, since knitted networks are representations of co-occurrence language networks~\cite{castro2019multiplex,stella2020forma,tohalino2018extractive,marinho2016authorship}.

The study reported in~\cite{Koponen2014} aimed at identifying key Physics concepts from students' representations of perceived similarity between distinct topics. The representation used in this work was a concept network, where nodes represent the concepts (in the sense of quantities, laws, models, or experiments), and edges represent similarities between these concepts, such as actions for determining a model or the realization of a experiment using some law ~\cite{Thagard1992}. The paper studies these concept networks using subgraph and communicability betweenness centrality. The most relevant concept networks were identified using an importance ranking coefficient, which is a normalized geometric mean of the considered centrality measurements. While this study does not relies on random walks to represent the acquired network, the concepts networks are used as examples of networks representing the knowledge acquired by students, according to unknown knowledge acquisition dynamics.

The study conducted in~\cite{Herrero2019} analyzed the properties of self-avoiding walks (SAW) in clustered scale-free networks. The study investigated how the number of SAWs changes as the desired walk length increases. The main result of the paper shows that, for scale-free networks with same average degree, there are more SAWs in clustered networks when compared to unclustered networks. This result suggests that the modular organization in the same topological family of networks may impact the discovery process in the network.


{
Differently from most of the works in the literature, here we analyze the  knowledge acquisition problem in terms of a generalist point of view. We analyze whether different network topologies and dynamics can lead to the same behavior in the observed learning curves. In other works, we analyze the behavior of learning curves by comparing, \emph{at the same time}, different configurations of network topology and agents dynamics.
%
}

\section{Methodology} \label{sec:methodology}

The main objective of this paper is to compare the efficiency of different walking strategy to discover new nodes in the network. We compare well known random walk strategies in different network topologies. Most importantly, we analyze the behavior of ``learning curves''  for each pair topology/dynamics in order to analyze whether different combinations of topology and random walks can lead to the same learning curve (and vice-versa).
The adopted methodology is illustrated in Figure \ref{fig:schematic} and summarized in the following steps:

\begin{figure}
\centering
\includegraphics[width=1.0\linewidth]{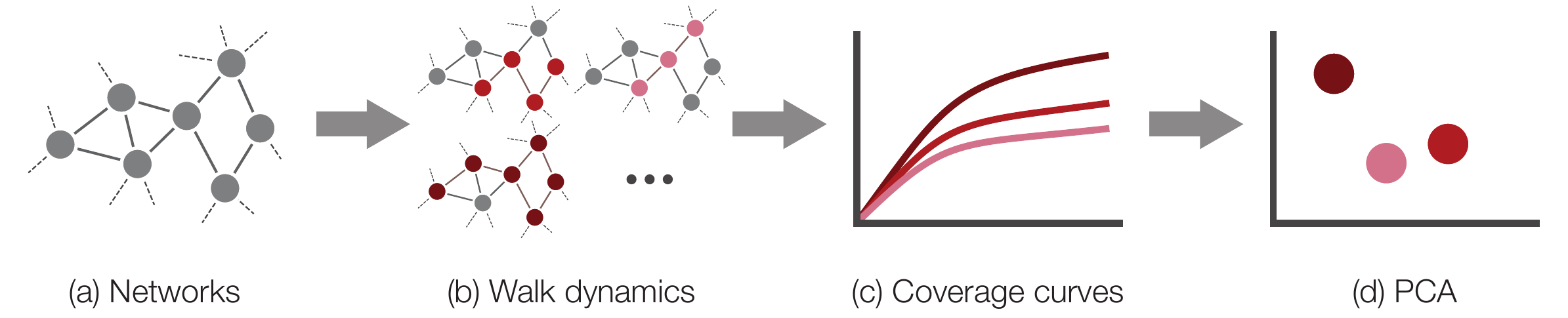}
\caption{Methodology employed to analysis the behavior of learning curves. In (a), we selected different network topologies. In (b), dynamics based on variations of random walks were considered to explore the networks. In (c), we obtain the learning curves describing how many nodes are discovered as the network is explored. Finally, in (d) each curve is mapped into a 2-dimensional space and similarities in the behavior of learning curves for different topologies and dynamics are analyzed. }
\label{fig:schematic}
\end{figure}

\begin{enumerate}

    \item \emph{Network topology}: we selected different network topologies. We have selected well-known network models reproducing the characteristics of real-world networks.
    A brief description of the adopted models is provided in Section \ref{sec:netop}.

    \item \emph{Network dynamics}: different ways to walk over the networks were considered, including dynamics based on traditional random walks and dynamics biased towards particular neighbor properties.  A brief description of the adopted network dynamics is provided in Section \ref{sec:dyn}.

    \item \emph{Learning curves}: For each pair of topology and dynamics, we obtain the learning curves. This learning curve describes how fast new nodes are discovered as the dynamics unfolds (see Section \ref{sec:lc}).

    \item \emph{Cluster analysis}:  in this phase, each learning curves are mapped into a vector. This is used to measure the similarity between two curves. Similar curves are then identified via cluster analysis.  This step is important to show that the behavior curve
    A brief description of this process is provided in Section \ref{sec:cls}.

\end{enumerate}


\subsection{Network topology} \label{sec:netop}

Artificial networks were built for each set of network models.
%
%
The following parameters were used to create the networks: number of nodes $(N) = \{500, 1000, 5000\}$ and average degree $(\langle k \rangle) = \{4,6,8,10\}$. We have worked with four well-known undirected network topology models:

\begin{itemize}

    \item \emph{Erd\H{o}s-Rényi (ER)}: this model generates small-world networks, adding the characteristic to have all the nodes with similar degrees, i.e., the probability of creating an edge is equally distributed among the nodes.

    \item \emph{Barabási-Albert (BA)}: this topology implements the scale-free model, inherent to many real networks. BA networks are characterized by a few hubs with a very high degree, while most nodes have small degrees.

    \item \emph{Waxman (WAX)}: this a traditional geographic model, which comprehends a set of nodes in a two dimensional space that incorporates new edges through an algorithm in which the probability decays exponentially as the distance between each pair of nodes grows.
    More specifically, the probability of two nodes to be linked is given by:
    \begin{equation}
        \pi_{ij} = a \exp ( d_{ij} / \beta ),
    \end{equation}
    where $a$ is a normalization factor, $d_{ij}$ is the geographic distance between nodes $v_i$ and $v_j$ and $\beta$ is a parameter that defines the connectivity of the network. 


    \item \emph{Modular Networks (LFR)}: networks with community structure were implemented using the methodology described in~\cite{Lancichinetti2008}. In this model, each community is represented as a scale-free network. In addition to the number of nodes and average degree, additional parameters can be considered to generate the networks. The main parameters describing this model are the number of communities ($n_C$), the minus exponent for the degree sequence ($t_1$), the minus exponent for the community size distribution ($t_2$), the maximum degree ($\max_k$), and the the mixing parameter ($\mu$),  which determines the fraction of edges linking distinct network communities.
    Here we used $n_C = 5$, $t_1 = 3$, $t_2 = 0$, $\mu = 0.20$. The maximum degree $\max_k$ were chosen so as to obtain networks with the desired average degree $\langle k \rangle$.



\end{itemize}

\begin{figure}[!hbtp]
\centering\includegraphics[width=0.82\linewidth]{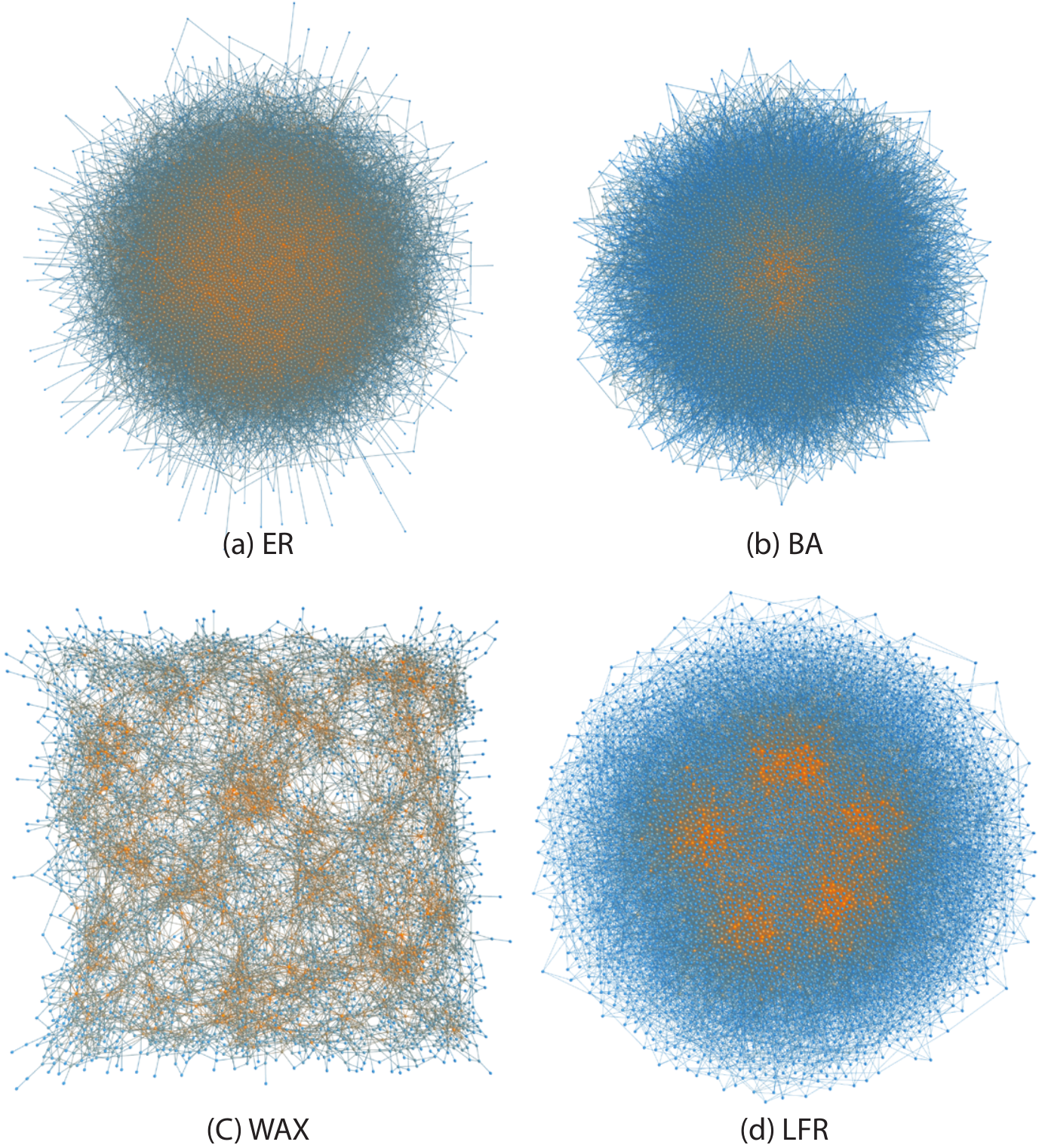}
\caption{Force-directed visualizations of the considered network models. Different colors correspond to different node degrees. The visualizations were generated using the \emph{Networks3d} software~\cite{SILVA2016}.}
\label{fig:viz}
\end{figure}

A visualization of the considered models for selected parameters is illustrated in Figure \ref{fig:viz}. The visualizations were generated using the \emph{Networks3d} software~\cite{SILVA2016}.
It is clear that for different models the nodes with highest degrees (orangish nodes) are distributed in different ways.

%



\subsection{Network dynamics} \label{sec:dyn}

In order to recover the symbols from these models we have worked with the following walk dynamics: traditional random walk (RW)~\cite{Lovasz1996}, random walk biased by degree (RWD)~\cite{Bonaventura2014}, random walked biased by the inverse of the degree (RWID)~\cite{Bonaventura2014}, and true self-avoiding walk (TSAW)~\cite{Kim2016,Amit1983}.
These walks have been widely employed to study the dynamics of learning curves in the last few years~\cite{ARRUDA2019,ARRUDA2017,lima2018dynamics}. The main differences among these walk dynamics are detailed below:

\begin{itemize}

    \item \emph{Traditional random walks}: the random walk dynamics is one of the most used in literature, and a very simple one. If the walker is at node $v_i$ and $\Gamma_{i}$ is the set of neighbors of $v_i$, all nodes in $\Gamma_{i}$ have the same probability to be chosen as next node in the walk. In other words, the probability of transition from $v_i$ to $v_j \in \Gamma_i$ is $p_{ij} = k_{i}^{-1}$.

    \item \emph{Degree-biased random walk}: in this walking dynamics, a higher probability of transition $p_{ij}$ is given to those neighbors with higher degrees. Mathematically, $p_{ij}$ is proportional to the degree $k_j$ of $v_j \in \Gamma_i$:
    \begin{equation}
        p_{ij} = \frac{k_{j}}{\sum_{l \in \Gamma_{i}} k_{l}}.
    \end{equation}
    In other words, the RWD dynamics always tries to explore the network by prioritizing visits to nodes with the highest number of neighbors.

    \item \emph{Low degree-biased random walk}:  a different variation of the traditional random walk is the walk biased towards the inverse of the degree. In this case, the probability of transition from $v_i$ to $v_j \in \Gamma_i$ is :
    \begin{equation}
        p_{ij} = \frac{k_{j}^{-1}}{\sum_{l \in \Gamma_{i}} k_{l}^{-1}}.
    \end{equation}
    Therefore, in this case, the walker tends to select nodes with low-degree in the next step of the random walk.

    \item \emph{True self-avoiding walk}: in a true self-avoiding walk dynamics, already visited nodes are avoided. This is achieved this by memorizing edges that have already been visited. The transition probability is computed as
    \begin{equation}
        p_{ij} = \frac{ e^{-\lambda f_{ij}}}{\sum_{l\, \in\, \Gamma_{i}} e^{-\lambda f_{il}}},
    \end{equation}
    where $f_{ij}$ is the frequency of visits to the edge linking nodes $v_i$ and $v_j$. The parameter $\lambda > 0$ corresponds to the exponential decay factor for which the probabilities decrease with the number visits. In this study, we use $\lambda = \ln 2$.

    The main advantage of this dynamics is that it tends to present a higher learning rate when many nodes have already been visited. When the walker is visiting a region with no visited nodes, this random walk behaves similarly to the RW dynamics. 

\end{itemize}


\subsection{Learning Curves} \label{sec:lc}

The measure used to characterize each dynamics is the so-called learning rate. This is an important property in network science and is related to many processes on complex networks, including knowledge acquisition, discovery processes, diffusion and spreading~\cite{da2007exploring}. For each pair of network and random walk dynamics, we considered 5,000 iterations (steps). Learning curves are then obtained as the fraction of the total number of \emph{different} nodes visited after a given number of steps.

The dynamics observed by visiting sequentially network nodes has an analogy with the process of generating written texts~\cite{ARRUDA2019}.  If we consider that, at each step, a symbol is generated  to represent that the current node has been visited, after 5,000 steps we have a sequence of symbols (i.e. a text) comprising 5,000 words. The learning curve can thus be seen as the vocabulary observed for a given text length. While in written texts the relationship between vocabulary size and text length is well described by the Heaps' Law~\cite{lu2010zipf}, the learning curve observed in network discovery processes tends to follow a different pattern~\cite{ARRUDA2017}.

\subsection{Principal Component Analysis} \label{sec:cls}


Here different learning curves are compared and similar learning curves is observed. To quantify the similarity between curves we represent each curve as $n$-dimensional vector, where the $i$-th position of the vector represents the fraction of nodes visited after the $i$-th step. Because such a representation of curves yields several strongly correlated features, we use Principal Component Analysis (PCA)~\cite{gewers2018principal} to remove possible correlations. In fact, as we shall show, two dimensions of the PCA analysis accounts for more than 95\% of the data variation.

After the learning curves are represented in a two-dimensional space, clusters can be identified. Because our objective is to analyze whether similar learning curves can be obtained with different topology/dynamics choices, the identification of clusters was performed via visual inspection. However, a scenario with several instances could also be analyzed by using traditional clustering algorithms~\cite{rodriguez2019clustering}.
%
%


\section{Results and discussion} \label{sec:results}

\begin{figure}[!hbtp]
\centering
\includegraphics[width=1.0\linewidth]{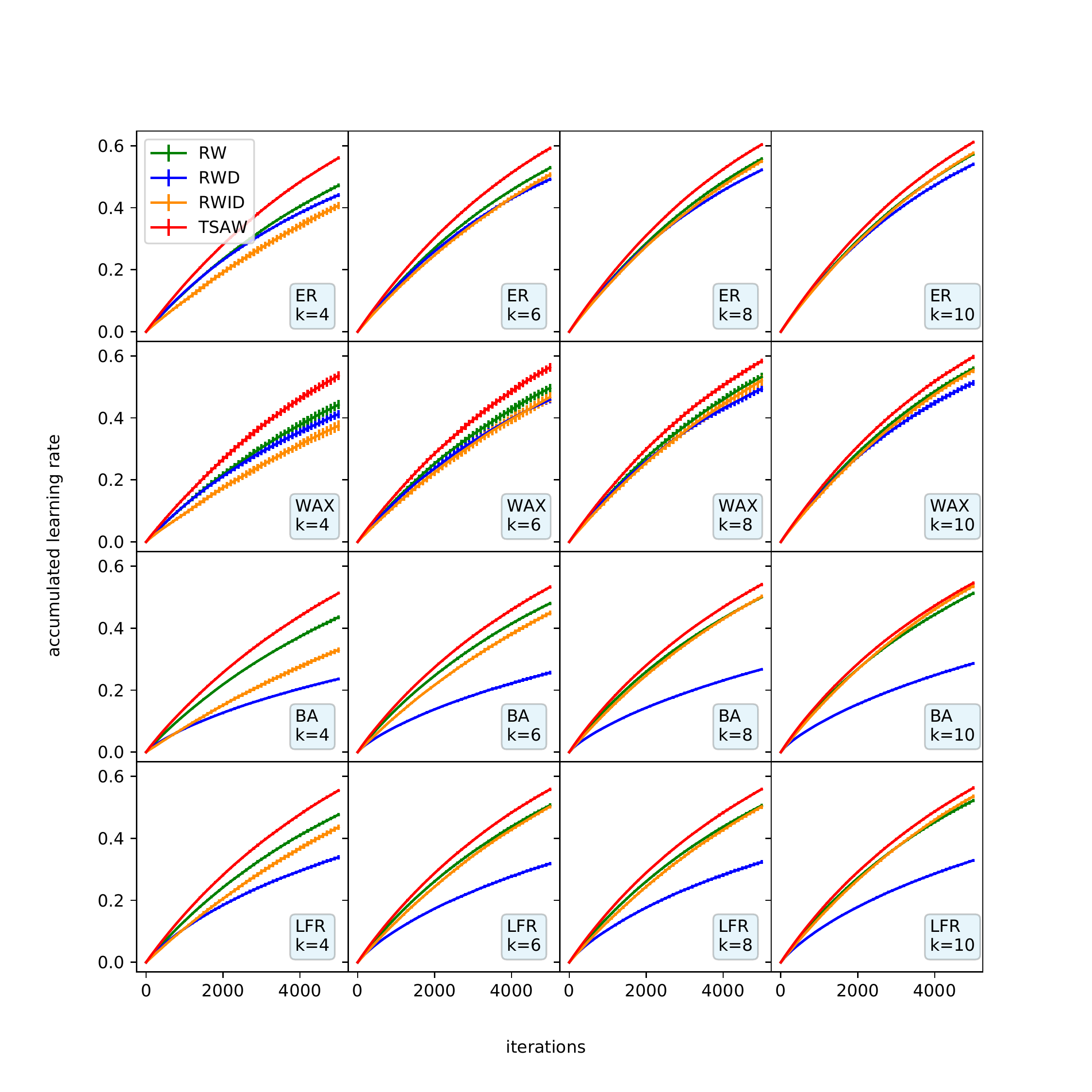}
\caption{Learning curves for $N=5,000$ nodes and the models ER, BA, WAX and LFR. Each row and column correspond to different network topologies and average degrees, respectively. }
\label{fig:allcurves}
\end{figure}

Our analyses take into account the exploration coverage over time for agents discovering knowledge in network models as they explore nodes through edges.
The first step is obtaining the learning curves for the considered pairs of dynamics (RW, TSAW, RWD, and RWID) and network models (ER, BA, WAX, and LFR models). For each network model setup, we generated $5$ networks and recorded the coverage curves for $50$ realizations of each dynamics. The starting position of each realization was drawn uniformly from the network nodes and for each configuration we computed the average and standard deviation of the coverage (learning) curves. The resulting curves are shown in Figure \ref{fig:allcurves}. Each row and column corresponds to different network models and average degree, respectively. The panels contain curves colored according to the considered dynamics.
%

An initial observation shows that the TSAW dynamics outperformed the other dynamics in all the experiments, corroborating previous studies in which TSAW was found to be among the most optimal stochastic walks~\cite{ARRUDA2017}. On the other hand, the RWD and RWID dynamic resulted in the worst performance among the considered configurations.  

All curves seem to present similar shapes but different growing speeds, with faster coverage as $\langle k\rangle$ increases, a behavior that is stronger for the RWD and RWID dynamics. In particular, for ER, the performance among the dynamics becomes substantially similar as the average degree increases. This indicates that the considered dynamics performs very similarly for denser networks. An exception to this rule is the RWD for the BA and LFR. In these cases, the performance of RWD gets slightly worse as network connectivity increases. This is probably related to the fact that a scale-free network (such as BA or LFR) allows the existence of extremely connected nodes in which a walker could get stuck given its preference to move to nodes with high degrees.

Another important aspect of the analysis is how the ranking of dynamics performance change amongs the experiments. In general, TSAW is followed by RW, except for the LFR and BA networks with high connectivity. In this case, RWID attains a second place. This reveals that, in these networks, avoiding hubs can be a good strategy to explore them more quickly. When the degree is lower, however, RWD performs better than RWID, indicating that, in this case, it is preferable to reach the hubs than avoiding them to attain better performance.

\begin{figure}[!hbtp]
\centering\includegraphics[width=1.0\linewidth]{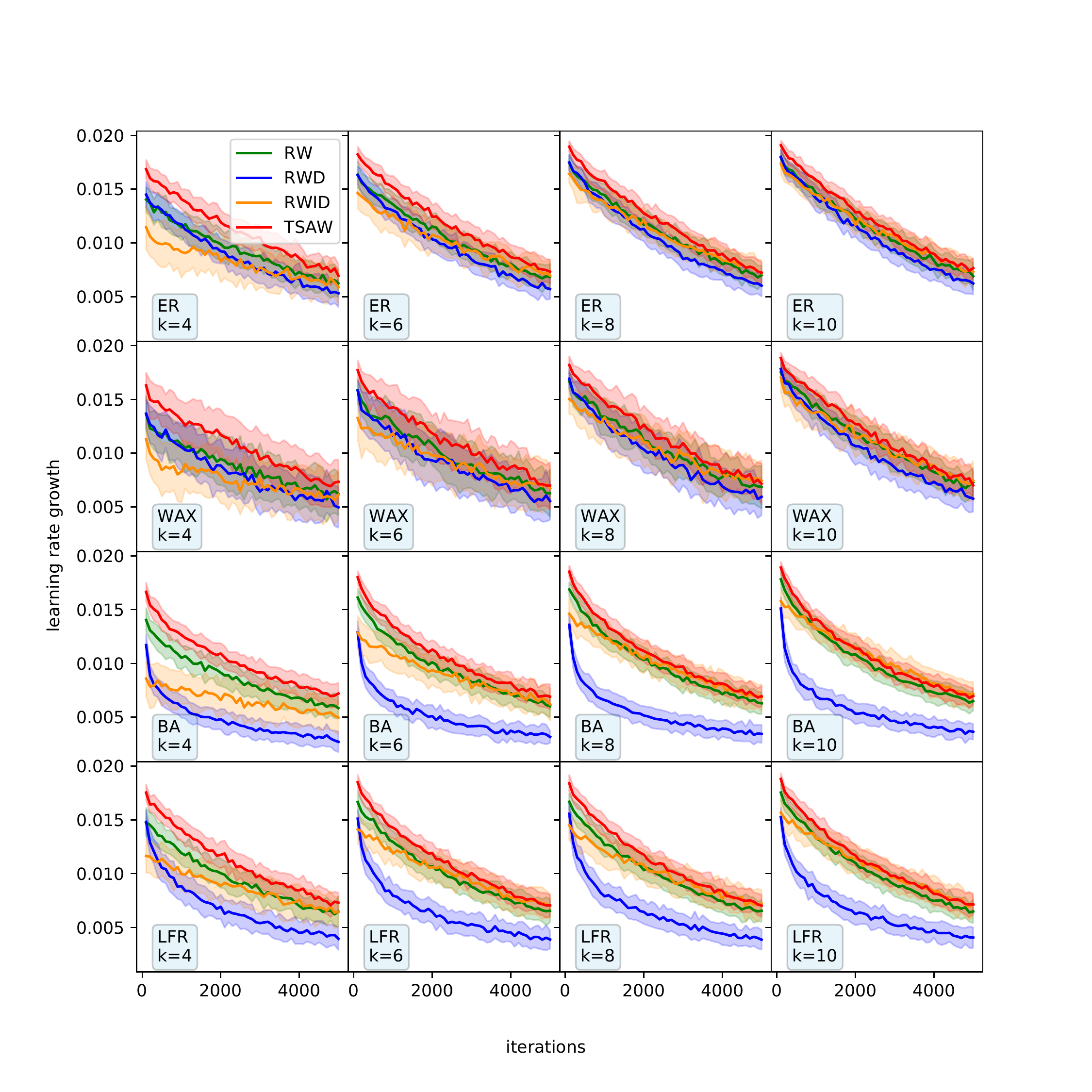}
\caption{Learning rates for the considered models and $N=5,000$. Each curve indicates the growth of the number of discovered nodes across the simulation epochs.
}
\label{fig:allcurvesdiffsstd}
\end{figure}

In addition to the previous analyses, we observe two distinct patterns for the behavior of the curves among the network models, one for ER and WAX, and another for BA and LFR. While these pairs do not necessarily display exactly the same behavior, the performance rankings of the dynamics within these pairs of models do not change much. We also analyzed the differences (or rates of growth) of the cumulative discovery curves. Figure~\ref{fig:allcurvesdiffsstd} shows the obtained rate curves for all the considered configurations. Both the ranks and other overall observations drawn from the cumulative curves can also be drawn for the rate curves.
%

%
To summarize the main characteristics of the obtained learning curves, we applied PCA as a way to reduce their dimension. For each experiment, we derive a set of $50$ features corresponding to the values of the learning rate curves (i.e., the derivatives shown in Figure~\ref{fig:allcurvesdiffsstd}) at epochs $100$ iterations apart (see Section \ref{sec:cls}).






%


\begin{figure}[!hbtp]
\centering\includegraphics[width=1.0\linewidth]{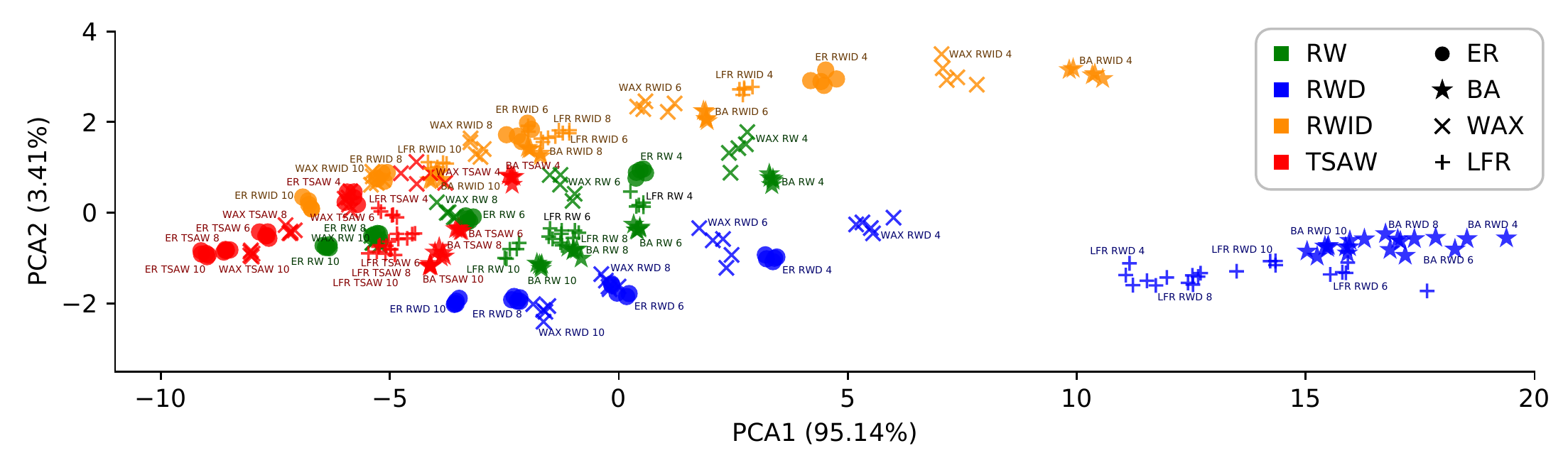}
\caption{PCA results for ER, BA, WAX, and LFR for $N=5,000$ nodes. Each instance represents a learning curve obtained for a specific pair of network topology and agent dynamics. Interestingly, in some cases, different combinations of topology/dynamics can lead to similar learning curves.}
\label{fig:pca5000_wide_labels_diffs}
\end{figure}

The obtained data projection, shown in Figure~\ref{fig:pca5000_wide_labels_diffs}, reveals that almost $100\%$ of the variance in the curves can be explained by only two components. In particular, the first component covers about $95.1\%$ of the variance. This outcome indicates a high correlation among the curves. At the positive extreme of the first principal component, we find a separated group corresponding to the curves obtained for RWD dynamics simulated on the BA and LFR networks. These correspond to the curves with worst performance among the considered experiments. The RWID curves spread across the PCA1 axis, revealing its diversified behavior with each curve depending on the network model and connectivity.

Along the negative segment of the first principal component, we observe a substantial overlap among the curves for different experiment configurations. This region corresponds to configurations of high node degree or simulated through the TSAW dynamics. Among the notable  overlapping configurations are ER and WAX. This is a surprising result, since they present very distinct characteristics in terms of global structure. At least three other regions are shared by different combinations of networks and dynamics. This includes those obtained from ER, WAX and LFR models when the dynamics are TSAW for LFR, and RW for the others. Another example are the RW curves for the BA, WAX, and ER. These results indicate that just by looking at the coverage performance curves it is not trivial to distinguish between network models and dynamics.

\begin{figure}[!hbtp]
  \centering
  \includegraphics[width=.7\linewidth]{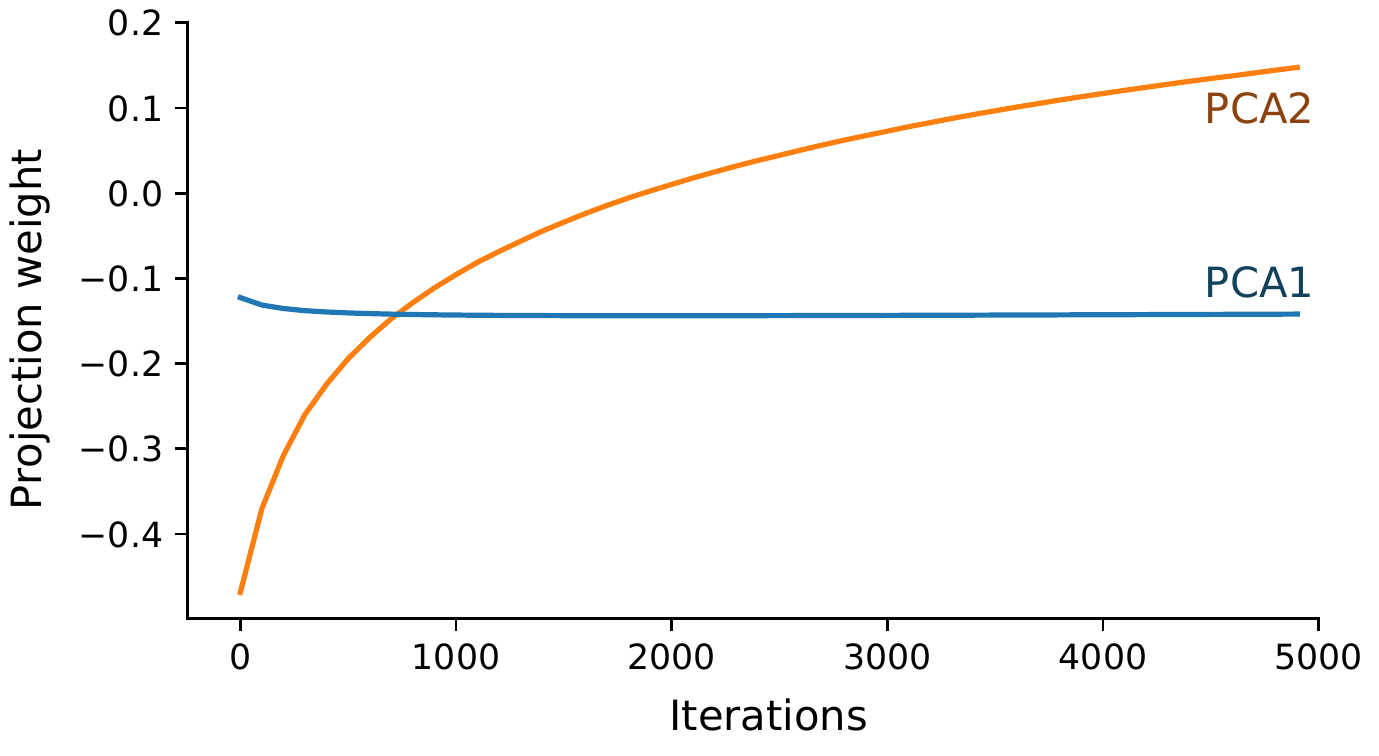}
\caption{Projection profiles of PCA1 and PCA2 axes along the original space.}
\label{fig:pca_axes}
\end{figure}

The profile of the PCA axes in the original space, shown in Figure~\ref{fig:pca_axes}, reveals that the first principal component (PCA1) is almost flat along the iterations. This indicates that all epochs are equally important for the principal component. Conversely, PCA2 seems to capture the difference of rates at the beginning and end of the curves. To further explore these aspects we plotted together all the averaged cumulative learning curves of the considered configurations colored by PCA1 and PCA2.
This result is shown in Figure~\ref{fig:pca_axes}. We note that PCA1 (a) indeed correspond to the inverse of total learning coverage, which is somewhat independent from the shape of the curves. A second order effect seems to be captured by PCA2 (b), corresponding to how fast the rates of the learning curves are increasing across the epochs. This becomes more clear when all the curves are aligned so that the starts and ends match, as shown in (c). Curves with low values of PCA2 tends to be more concave (presenting high curvature) and vice-versa. All in all, PCA1 corresponds to the average learning speed, while PCA2 seems to be related to the acceleration of the curves.

\begin{figure}[!hbtp]
  \centering
  \includegraphics[width=0.95\linewidth]{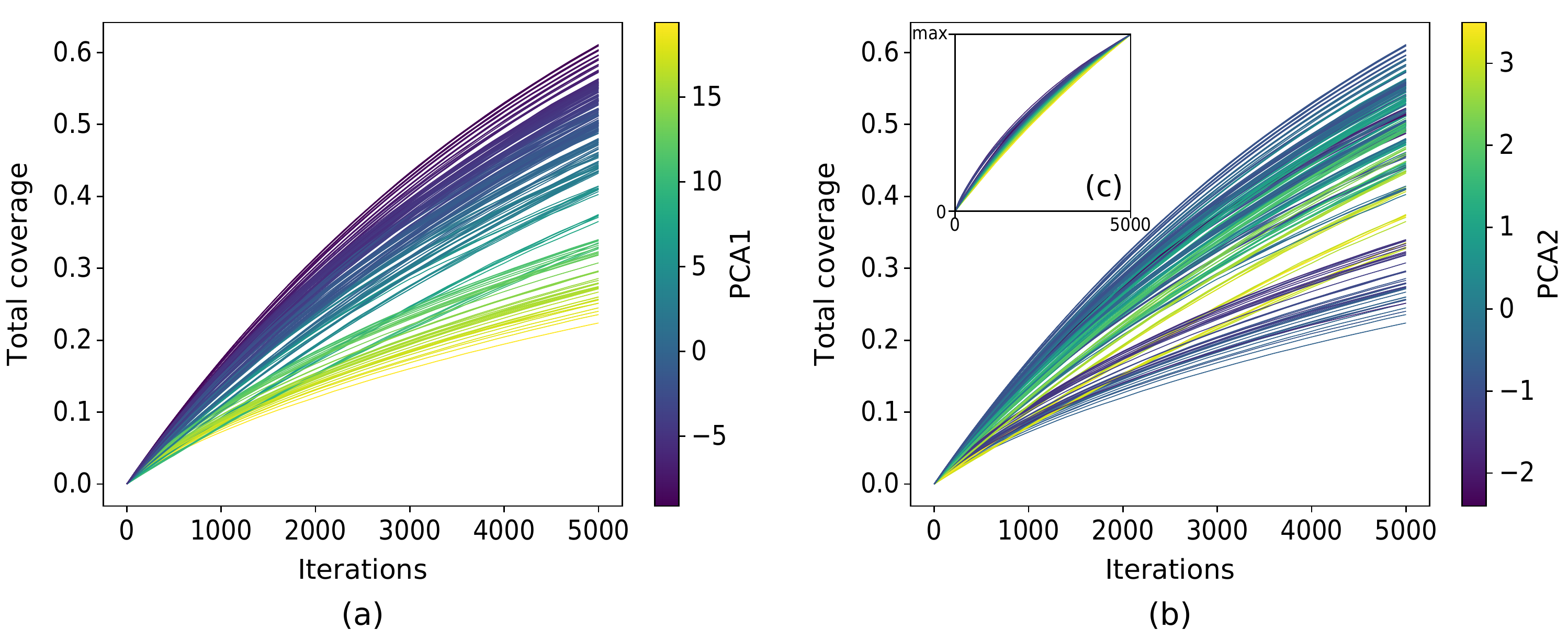}
\caption{Averaged learning curves for all the considered configurations. The color of each curve indicates the PCA1 (a) or PCA2 (b). The insight (c) shows all the curves normalized by their respective maximum value.}
\label{fig:pca_axis}
\end{figure}


\section{Conclusion}

With many real-world phenomena being modeled and represented as sequences, one way to characterize their respective complex system is by separating the dynamics encoding the sequences from their underlying state space. In this context, a certain stochastic walk dynamics acts as the encoder while a complex network can be used to represent the state space. While this framework has been used to model several real-world problems, no systematic analysis of the relationships among these three aspects of the systems exists in the literature.

In this paper, we performed a systematic analysis of the behavior of different dynamics in well-known network topologies. Whenever a dynamics (or exploration strategy) is performed on a network, one obtains a sequence of visited nodes. We aimed at studying how both topology and network dynamics affects the observed sequence of visited nodes. Here we focused in one property of the sequences, the total number of different visited nodes. This property has many applications in network science, and is oftentimes related to the process of knowledge acquisition~\cite{ARRUDA2017,ARRUDA2019}. In a semantic network, for example, each visited node can be considered as a new learned concept.

We adopted a framework to study the behavior of learning curves. For each combination of network topology and dynamics, we obtained the corresponding learning curves. Then, each learning curve was mapped into a two-dimensional space via Principal Component Analysis. This allowed us to compare curves in a more systematic way, with the advantage of removing correlations while keeping the variability of the original learning curves space.

Several interesting results have been found with our approach. Overall we found that true self avoiding walks outperformed all other dynamics, while the variations of random walks biased towards high or low degree displayed the worst learning curve performances. Despite such differences in performance, we found that all learning  curves presented similar shapes. A further investigation of growth rates (i.e. the derivatives) of learning curves revealed that no additional information can be obtained from such an analysis. This means that the learning curves are sufficient to discriminate different network topologies and dynamics.

The Principal Component Analysis confirmed that, despite distinct performances, all curves shapes are similar. This could be confirmed by the fact that curves could be mapped into a two-dimensional space virtually without any lost in the original data variation. Surprisingly, the first component accounted for 95\% of the original variation. The visualization provided by PCA allowed us to observe some interesting patterns. Some regions were found to share different combinations of topologies and dynamics. For example, similar learning curves were found in ER and WAX, showing that the same behavior can be obtained even in very distinct network topologies. The PCA visualization also revealed the variability of learning curves with different topologies. While RWD and RWID were found to be very dependent upon topology, learning curves obtained with TSAW dynamics were found to be much less sensitive to distinct network topologies.

The ambiguity of the behavior of learning curves observed in the PCA space can be useful in practical scenarios. For example, in a knowledge acquisition scenario, the network topology can represent how concepts are linked to each other, while the chosen dynamics can be interpreted as the methodology used to cover the concepts being taught. In such educational scenario, our results suggest that one can be able to deliver the same learning experience by adopting completely different knowledge organization (i.e. network topology) and teaching sequence (i.e. network dynamics).

Our results show that when one uses learning curves to describe sequences of visited nodes ambiguous behaviors may arise. In other words, sequences with similar behavior can be observed from distinct pairs of topology/dynamics. This result suggests that the reconstruction of the processes underlying network construction and topology cannot rely only on learning curves as descriptive features of sequences. For this reason, in future works, we intend to study additional sequence features to identify a minimum set of sequence descriptors that are able to discriminate both the topology and dynamics generating the observed sequence. Because sequences are used to construct embeddings, further studies can analyze if similar embeddings can be obtained from distinct topologies and walks.

\bibliographystyle{ieeetr}
\bibliographystyle{abbrv}







\end{document}